\documentstyle{article}
\begin{document}

\title{The Twilight of the Scientific Age}

\author{Mart\'\i n L\'opez Corredoira\\
Instituto de Astrof\'\i sica de Canarias,
E-38200 La Laguna, Tenerife, Spain\\
Departamento de Astrof\'\i sica, Univ. La Laguna,
E-38206 Tenerife, Spain\\
E-mail: martinlc@iac.es}

\maketitle

{\bf \large ABSTRACT}

This brief article presents the introduction and draft of the fundamental ideas developed at length in the book of the same title, which gives a challenging point of view about science and its history/philosophy/so\-cio\-lo\-gy. Science is in decline. After centuries of great achievements, the exhaustion of new forms and fatigue have reached our culture in all of its manifestations including the pure sciences. Our society is saturated with knowledge which does not offer people any sense in their lives. There is a loss of ideals in the search for great truths and a shift towards an anodyne specialized industry.


\vspace{5mm}

\begin{center}
----------------------------------
\end{center}

\section{Introduction}

\begin{quotation}
``Yes, yes, I see it; a huge social activity, a powerful civilization, a lot of science, a lot of art, a lot of industry, a lot of morality, and then, when we have filled the world with industrial wonders, with large factories, with paths, with museums, with libraries, we will fall down exhausted near all this, and it will be, for whom? Was man made for science or science made for man?'' (Miguel de Unamuno --- {\it Tragic Sense of Life} [1913]; original in Spanish; Translation into English made by myself)
\end{quotation}

This quotation reflects quite accurately the main theme of the present pages. Read it carefully, twice or thrice, think about it for some minutes, and then begin to read the following pages as a musical piece whose leitmotiv is Unamuno’s assertion. Just a few minutes, or even seconds, may be enough for the reader to realize the most important message that I want to develop, and its connection with the title of this article (very abridged version of the book of the same title: L\'opez Corredoira, 2013). The idea is simple: our era of science is declining because our society is becoming saturated with knowledge which does not offer people any sense of their lives. Nevertheless, in spite of the simplicity of this idea, its meaning can be articulated in a much richer way than through one sentence, as in the case of a music which develops variations on a folk melody.

There are several reasons to write about this topic. First of all, because I feel that things are not as they seem, and the apparent success of scientific research in our societies, announced with a lot of ballyhoo by the mass media, does not reflect the real state of things. Also, because the few individuals who talk about the end of science, do so from relativistic or antiscientific points of view, not believing that science really talks about reality, or they relate the scientific twilight to the limits of knowledge. However, there is a lack of works which question the sense itself of the pursuit of the truth among present-day thinkers. Of course, there are many humanistic approaches which simply ignore science, but ignoring is not the same as considering its sense or lack of sense. There are many well-prepared scientists or journalists who move in the world of science and consider it in their interactions with the rest of society, but usually they focus too much on the scientific and technical details and do not go deeply enough into existentialist or subjective approaches. A wider vision of both worlds, those of the humanities and science, is necessary to undertake the task. I feel I am able to offer something of this sort, given my experience as both scientist and philosopher. It is not a matter of virtuosity in either scientific knowledge or other areas but a matter of being able to integrate a global view of the fate of our societies. Normally specialists are too focused in their narrow or biased views to offer a global analysis and feeling.

When we talk about the sense of something, we cannot undertake a pure analysis in objective terms as in a scientific study. The professional activities on those who dedicate their lives to natural or social sciences usually overlook the fact that, after all, human beings do not move because of {\it reasons} but because of {\it emotions}. As psychoanalysis claims, most of our actions are determined by unconscious impulses. And science itself is not an exception: It is made by men whose motivations stem from factors other than a mere pursuit of knowledge. We are not machines, we are not gods; we are just animals, very peculiar animals and very intelligent and curious, that make scientific enterprises work, but subject to multiple internal and external conditions. 

Societies as a whole are also sensitive to motivation. As a matter of fact, not all societies developed science. And, as it is known, even civilizations which developed that world-view and that methodology of observing phenomena can decline and lose their interest for continuing the scientific activity. That happened in Western Christian countries in the Middle Ages. Were the Middle Ages a dark age? Possibly, from some intellectual points of view, but it was not the end of civilization. It was an era with plenty of resources to create magnificent things, such as cathedrals. There were means to carry out great advances in many areas. Christianity was not intellectually underdeveloped with respect to Muslim countries, and basic knowledge of Greek science was also present; however, with very few significant exceptions, there was not a great development of sciences in Christian Europe during nearly the ten centuries of the Middle Ages. Why? Maybe because people were not motivated enough to think about nature. Surely, religious context had something to do with this, and the philosophy associated with religion which was ordered to follow faith above all. But possibly this is not the full explanation: The great revival of science in the Renaissance took place within similar religious creeds; also, the Muslim religion was not so different to Christianity and allowed in the Middle Ages a higher development of sciences, declining later when science in Christian countries began to dominate.  

In our era, the conditions are very different to the Middle Ages. Nonetheless, in a not very far future, societies embroiled in a lot of survival problems (overpopulation, lack of energy resources, economical crises, global warming and other ecological disasters, wars, plagues, etc.) may begin to see research as an activity that is not profitable enough and may abandon pure science research. At the beginning, people will trust scientists to solve all their problems, as it happens now, but they will realize that science cannot satisfy all those expectations, and that the returns of hyper-millionaire investments are smaller and smaller, nations will reduce more and more the titanic economic efforts necessary to produce some tiny advances in our sciences, to a point where scientists will say that they cannot continue their activity with such small budgets; consequently, the research centres will begin to close, one after another. Is this the prophecy I want to develop? No, I do not want to talk about prophecies. The future is uncertain and what I have described is only one possibility among many others. I want to speak about our present society, and the trends that can be observed now. 

Normally, throughout History, thoughts occur in advance of acts. What we are observing around us now are the effects of an ideology which was in some minds many decades or centuries ago. There is a slow inertia in societies which makes them move at the rhythm of impulses that originated some generations back. Geniuses are in advance of their time; what is famous at any moment is representative of a tradition of old, worn-out ideas. Religions gained their maximum power and influence a long time after they were developed: Popes and priests in the Renaissance, embedded in corruption and malpractice, with much less idealism than the conceivers of the religious ideas, were dominant in a time in which the most important creators were pointing to other directions. Today, science and some of its priests occupy an important status in our society, and gargantuan amounts of money support them. A superficial view may lead us to think that we live in the golden age of science but the fact is that the present-day results of science are mostly mean, unimportant, or just technical applications of ideas conceived in the past. Science is living on its private income. 

My interest is to lift the curtains behind the stage of science, and see what is going on in the engine room. If we want to ascertain which will be the next performance on the stage, it is better to see the organization from inside rather than just assisting with the show. In any case, I insist, I am not a prophet and it is not my mission to say how the future will be. Also, it is not my mission to give a report of all the observed trends and ideas around the world of science. What I will offer is my personal view, not necessarily reflecting the views of all conformist and non-conformist present-day thinkers.

The {\it leitmotiv} is a simple melody. Its harmonization with other melodies and rhythms and the orchestration which integrates all the voices is a more complex thing. As in Wagner’s operas, we pursue an infinite melody: A continuous flow where the main melody gets lost among instrumental and human voices. The question of the sense or non-sense of the human endeavour called science must take into account many circumstances. The exhausting of important ideas to explore, the limit of knowledge, is part of the matter. The excess of information is another part. But there are more questions to tackle. The question about the sense of all this stems from those different sources, like a river that takes water from its tributaries, and also from the need for introspective reflection. From time to time, it becomes necessary to go away from the river and contemplate it from the shore. Where does the river go? To the sea, we shall answer. And what for? Is it to achieve Truth? Is it to dominate Nature? What for? For whom? Was man made for science or science made for man?

Thinking about the role of science in present-day society is thinking about the past and the future of humanity. Human beings must question from time to time all their principles and their usual ways of life. There is nothing sacred and untouchable. The missions that science had in the past have been totally accomplished, or almost totally. Now, it is time to reflect anew on our society for the future, not only science but also many other activities or concepts: Art, religions/sects, History, universities, economic systems, political systems, human rights, etc. Very few things are permanent, and all of them are biological, such as taking food and water, sleeping, having sex, etc. All cultural things are subject to change; there is nothing eternal in them. From an anthropological point of view, all the characteristics of our civilization are simple features of the human specie in a given period of time and a given geographical localization. Certainly, the success of Western culture, with the subsequent annihilation of other cultures, has expanded the geographical location of our civilization to the whole planet, and this might lead us to think that our concepts, such as the so-called human rights, are absolute and universal. A mirage, an illusion! We just live our moment of glory, such as those of many empires which have absorbed great portions of land. The Roman Empire and the Egyptian civilization were greater than us; they lasted longer periods of time, dominating relatively large portions of land for that era. They were perhaps as proud as we are of our Western culture but they eventually declined. Now, it makes no sense to us to bury and embalm the pharaohs under pyramids. Possibly, future civilizations will not see any sense in building huge particle accelerators or telescopes. 

You may think that the pharaohs were wrong in their belief that they could preserve life after death, whereas we are right in our scientific truths. I agree. I am not a stupid cultural relativist: Of course, atoms exist and they are constituted by subatomic particles; of course, galaxies and stars exist. But the question is not about the truth of scientific assertions but about the place these truths occupy in our lives as human beings. In the Egyptian civilization or in our civilization, we are moved by our beliefs about what are the high values for our lives. The pharaohs believed that the great architectonic efforts of their people were worth it because that would allow them to be closer to eternity after death, and to show their status on earth too. Scientists believe that dedicating their lives to scrutinizing the laws of nature and making a complete catalogue of all the existing forms of matter, either inert or alive, will bring them closer to something eternal: truth; and make some profit on earth too... But then a question like that of Unamuno arises: ``when we have filled the world with industrial wonders, with large factories, with paths, with museums, with libraries, we will fall down exhausted near all this, and it will be, for whom?'' Is not it like the child of the tale {\it The Emperor’s New Clothes} that wakes us up from our dreams?

Behind the search for something permanent in our lives, something eternal, something absolute, there is most likely some fear of death. Death is an unavoidable topic if we are going to talk about the sense of some activity for our lives, or the sense of life itself, because precisely our certainty of the finiteness—and indeed very short compared to our aspirations—of our lives pricks our need to search for a sense. We waste our time: we will never find any sense in terms of eternity, but culture is fed mostly because of these aspirations, so the belief is not a bad business at all. Indeed, culture might be understood as the attempts of a civilization to alleviate the tension of the uncertainty which produces our certainty that we are going to die. From this psychological point of view, science is just one of the performances of this tension on stage among many possibilities. 

History shows us many dawns and twilights in the different facets of human beings. Looking at the past we can date and understand the reasons for the birth of science. We do not know when its twilight will occur, but the reasons for it are already in the air: after a very hot summer always come the season for the drop of leaves. 

\section{Some Highlights in the History of Natural Sciences}

If we want to understand why science is declining, we must first understand how it became so important. We must look at some of its advances to realize how important science has been for our culture, otherwise any attempt to talk about the twilight of the scientific era sounds like an anti-scientific complaint, and that is not my purpose. 

There are some instances when science becomes very important, affecting not only specific subjects but also general world views which change our philosophy, religious beliefs, etc. This is the case with the Copernican revolution. That a creature on the earth is able to understand the position and motions of the planet, despite the appearances to the contrary, may be called properly ``intelligence''. The search for truth in science is a heroic enterprise, and the greater the difficulty in reaching it, the more value it has. In the revolution which bears his name, Nicolaus Copernicus was the great hero, although not the only one, to judging from the number of people before and after him who worked on the same idea. 

There have been many phenomena around us whose logic was not understood and, in the era of rationalism and the subsequent century of enlightenment, science would offer explanations about them in a positive way, distinct from metaphysical speculations. Understanding the mechanics of nature was a dream pursued by human reason during many centuries and, with Newton’s system, the great triumph of modern physics was reached:\footnote{The term {\it physics} is generally used after 1850, instead of {\it natural philosophy}.}  a set of principles and laws which allow us to understand and to calculate motion in terms of kinematical principles and known forces, a triumph also in the understanding of gravitation as well as of celestial mechanics. 

The most important contribution to the understanding of the mechanics of nature and its forces after Newton involved another synthesis of knowledge, this time on the subject of electromagnetism, produced by James Clerk Maxwell, achieving the unification of electricity, magnetism and optics. Understanding the mechanics of nature would mean understanding how nature worked in all its aspects. That was the major aim of science, a dream pursued for a long time. This view, as seen by Huxley (1895) and many others at the end of the nineteenth century, is called ``reductionism''. Most present-day professional philosophers are against reductionism. They see it as a major threat to metaphysical speculation, and they prefer to adopt either a non-naturalist position or claim a mysterious emergence of irreducible properties in nature. Most present-day scientist however do not even talk about reductionism but take it as an undeniable fact.

The development of atomic theory is one very important achievement of science and it is one of the symbols of solid knowledge. And, since everything in nature is matter constituted by atoms, the understanding of this fact is of capital importance. Practically, all sciences are fed by atomic theory. In the last decades, particle physicists have made great efforts to learn more about the structure of matter, but with less spectacular results; the only impressive thing about particle physics nowadays is the gargantuan amount of money which it is able to consume in a short time. 

In my opinion, the most important of all scientific revolutions is the theory of evolution. Why? Because this revolution affects not only our concept of nature, or our concept of matter, but also our understanding of ourselves. It is the answer to questions such as: Who are we? Where do we come from? Its importance surpasses the limits of science or pure knowledge. Of course, the debate about religious beliefs was and is still important, but it is more than that. It is perhaps the scientific theory with the greatest relationship to the humanities in general, philosophy, anthropology, history (prehistory), etc. It is a pinnacle in human wisdom, perhaps the highest. Along with the theory of evolution and genetics, the other great pieces in the puzzle to understand the nature of life were biochemistry and molecular biology, which made their greatest developments during the twentieth century. Their purpose, to explain life in terms of physics and chemistry, is a great reductionist enterprise. 

These examples of scientific highlights are enough to recognize the important role of natural sciences in the past for our world. And, however, the resources dedicated to research until a century ago are negligible in comparison with the resources devoted to it in our own society now. It is a datum to take into account in our reflection about the sense of promoting fundamental research: increasing expenses with diminishing returns. 

We may wonder why science in the past was so cheap, or why it is now so expensive. There are three reasons: 
\begin{enumerate}
\item The ratio of scientists to general population in the past was lower than now. The number of scientists and engineers dedicated to R\&D in the present day is 5.8 million (OECD data, 2006, excluding India); that is, on average, around a thousandth part of the world population. In developed countries the ratio is much higher: for instance, in U.S. it is around one in two hundred. In the whole of the nineteenth century, the number of persons around the world who published at least one paper in a scientific journal, either pure or applied science, was approximately 115,000 (Gascoigne, 1992), although most of them only published one paper and did not dedicate their lives to science. The number of people who lived through the hundred years of the nineteenth century was around 3 billion, so a ratio of scientists to general population of $\sim 4\times 10^{-5}$ stems from that census, even including people who only produced one paper, is lower than the current one. The number of scientists in the seventeenth century is roughly ten times lower than in the nineteenth century (from the analysis by Gascoigne, 1992, of scientists who appear in the Historical Catalogue of Scientists and Scientific Books) while the population was roughly half that of the nineteenth century, so the ratio in the seventeenth century would be circa $10^{-5}$.

\item The instruments needed by empirical scientists in the past were much more rudimentary and easy to build, and consequently cheaper. That expenses for instruments are much higher now than in the past is quite obvious. Certainly, the telescope which Galileo built was much cheaper than the 39m E-ELT telescope to be built by {\it European Southern Observatory} (ESO) in Chile with a planned cost of one billion Euros; and space telescopes are even more expensive.

\item Most of the scientists in the past did not receive a salary for their research. In most cases they even paid their own research expenses, and they did not have as many paid vacations disguised as “conferences” as they do now. The funding of science has changed a lot nowadays with respect to the past. The professionalization of research is actually a very recent thing. 
\end{enumerate}

\section{Institutionalization of Science and its New Socioeconomic Conditions}

Science has gained recognition from society, and it is nowadays one of the centres of power which pulls the strings of our society. It is a new church. Many times, philosophers have compared science with religion. I do not think that comparison is appropriate. Science is an activity very different from religion, and its concepts have an empirical basis which is far beyond the beliefs of religion. Nevertheless, from a sociological point of view, just looking at the social organization, there are certainly some similarities. 

The project’s main researchers are leaders of a group comprising several Ph.D. students, several postdocs and, perhaps, some senior scientists of lower status. This main researcher is usually a kind of commercial manager. They begin their careers as scientists, but they become administrators or politicians of science. The situation is well described by Gillies (2008, ch. 8): ``Academics typically start with great enthusiasm for research, but, after a number of years working at research, they often become rather bored with it. They may have run out of ideas. They may have come to realise that their youthful hopes of becoming the next Einstein were an illusion, while the reality is that there are quite a number of young researchers doing better than they are. In these circumstances the sensible move is into administration and management where a tempting career ladder stretches before them''.

The snowball effect, also called the Matthew\footnote{Merton (1968) gave it the name “Matthew effect” from the Gospel of Matthew (25:29) which says: ``For everyone who has will be given more and he will have an abundance. Whoever does not have, even what he has will be taken from him''.} effect (Merton, 1968), is present to a certain extent in the social dynamics of science, especially in the most speculative areas. It is a feedback loop: the more successful a line of research is, the more money and scientists are dedicated to working on it, and the greater the number of experiments on observations that can be explained {\it ad hoc}, such as in Ptolemaic geocentric astronomy; this leads to the theory being considered more successful. 

In some cases, the system supports conservative views, but there are also cases of speculative lines of research that have been converted into large enterprises. For instance, in theoretical physics, string theory has absorbed a lot of people and funds, as well as marginalising and deprecating other approaches to the same problems (Luminet, 2008). Smolin (2006) thinks that string theory is not only speculative but the conclusions are circular, the concepts are arbitrary and the hierarchical structure of this scientific community is quite outlandish. The Nobel Prize winner in Physics, Sheldon Glashow, wonders whether string theory is not more appropriate for an Institute of Mathematics or even a Faculty of Theology rather than to an Institute of Physics (Unzicker 2010, ch. 14). Unzicker (2010, ch. 14) considers physicists working in that theory as being like a sect or mafia. Another case is the search for supersymmetric particles in dark matter, which occupies more than thousand people at CERN. And what happens when, after a long period of search, when huge amounts of money have been consumed, the experiments or observations do not find any evidence in favour of these theories? Then the groups claim that we must carry out exploration at higher energies and they ask more money.

 As is well known, control of communications and practice of power are closely related. Thus the system, far from allowing free publication of results among professionals, works hand in hand with censorship. Theoretically, this control is presented as a quality filter but its functions are frequently extended to the control of power. Those researchers who want to publish in scientific journals are subject to the dictates of the chosen referee and the journal editors, who will say whether the paper is accepted or not: this is the peer review system. There is plenty of evidence of bias in support of papers confirming the currently accepted viewpoint and in favour of established researchers (Armstrong, 1997). Nepotism (friendship networks) is also common (Wenneras \& Wold, 1997; Thurner \& Hanel, 2010). Nonetheless, actually, the main problem is not direct censorship itself, but the screening action of the massive overproduction of papers, with millions of scientists producing millions of papers every year, the reading of which cannot be undertaken by even the most hardworking of readers. This means that, once the obstacle of direct censorship in the journals is removed, the researcher who tries out new ideas will have to fight with indirect censorship: the super-production of papers that conceal what is not of interest to the system. Propaganda is the key element in a paper becoming known. For this, the leading specialists again have the advantage, because they control most of the strings which move the publicity machinery; they have the appropriate contacts, they write reviews (summaries of scientific discoveries within a field), they organize congresses and give talks as invited speakers. Moreover, the reproduction of standard ideas is more acceptable because many people are interested in them, whereas the diffusion of new ideas is of interest only to their creators. 

This is not something new, it has happened all throughout history. The new thing is the institutionalization and bureaucratization of this process.

\section{The Decline of Science}

Some of the problems of science nowadays, which constitute some symptoms of the decline of our culture, are:
\begin{enumerate} 
\item Society is drowned in huge amounts of knowledge, most of it being about things of little importance for our cosmic vision, or producing no advances in the basic fundamentals of pure science, only technical applications or secondary details.

\item In the few fields where some important aspects of unsolved questions have arisen, powerful groups control the flow of information and push toward consensus truths rather than having objective discussions within a scientific methodology; it gives few guarantees that we are obtaining solid new truths about nature. 

\item Individual creativity is condemned to disappear in favour of big corporations of administrators and politicians of science specialized in searching ways to get money from States in megaprojects with increasing costs and diminishing returns.
\end{enumerate}

We can use one adjective to describe the status of science at present and in the near-future: {\it decadent}. It is only a subjective perception. Possibly other people will think the opposite thing, that we live in a golden age of science. There are plenty of reasons in favour of the first thing (see in López Corredoira, 2013 many examples of malpractices). Rather than a question of pure argument, it is a also question of sensitivity, of being able to perceive the sense or nonsense of the major enterprises which are nowadays called science from a human point of view. The quantity of publications, the quantity of big instruments and the technology created, the number of jobs created in research, the accurate control of our science in comparison with past times, etc. might be arguments to show that science is presently living in a wonderful epoch. However, I would reply, the spirit of science is being lost. And how do you measure the quantity of spirit? No, it is not a measurable quantity; forget about creating a new {\it scientometric} method to determine the amount of scientific spirit. It is a question of sensitivity: just look around; just talk with some leading scientists and observe their lives, their work. Technocracy is replacing the joy of scientific creativity. 

The same thing could be said about poetry: Do you think we live in a golden age of poetry now because a huge number of poems can be found on the internet, there are a lot of poetry competitions with hundreds of participants, and there are many poetry clubs, etc.? No, the spirit of poetry is nothing to do with that. What then is? If you cannot find an answer yourself, it is because you are not sensitive enough to poetry. Something similar happens with sciences. It is necessary to be sensitive to scientific thought in order to appreciate its boom or decadence.

The fact that science has become a big enterprise, consuming huge amounts of state funding has made it more vulnerable to becoming politicized and subject to the social values of the masses rather than the values of a thinking elite. Not only are scientists at the service of mediocre programs of research devised by mediocre scientists who dedicate more time to bureaucracy and getting funds than to thinking about science; we have even reached the incredible situation that ordinary people without much idea about science are being asked to propose topics for the future direction of science. The argument is that people pay their taxes and scientists use part of these taxes to do their science, so people have the right to choose which projects the public money is invested in. See, for instance, the web page 
http://www.reto2030.eu/, where people may vote for which scientific projects should be financed before the year 2030. Not only mediocre scientists choose, now non-scientists also choose. 

The problem for scientific institutions will probably come when its influence over society is reduced and when the resources that science consumes begin to diminish. One possible reason for stopping the expansion of scientific investment and causing its collapse is that science will reach the maximum expense that a society can afford. Scientific institutions follow the structure of capitalism, so they must continuously grow. Experimental science becomes more and more expensive with time, and science has opted for this way of no return, going always for an increase in funds. When a limit is reached at which the investment in science can grow no more, a crisis will become unavoidable. Nowadays, the richest countries invest around 3\% of GDP in research and development, from which 20\% is for pure sciences, a ratio much higher than in the past, both in absolute and relative terms, and that has grown continuously in the last few decades, with some small fluctuations. Possibly this is already close to the asymptotic limit in terms of the relative ratio of money that a society can afford, so a crisis may be not very far away. The crisis will also depend on circumstances in society: if the GDP of developed countries grows through inflation, even at a constant ratio it will result in an increase of investment in science; the exponential fast growth of the last few decades cannot be sustained but at least a slow growth may delay the death of science. The effect will not be immediate. Possibly, many centres will continue for some decades with a decreasing budget, but eventually they will recognize that no advances can be made with small budgets; even less than the few ones obtained with the huge budgets they have got nowadays. Therefore, research centres will begin to close, one after another. 

A crisis in the business of science, a crisis of no return, may happen, and a dark age in the advance of scientific knowledge might arise. This will not happen very fast but will be a slow process, possibly lasting several generations, and this decline will not only affect science but the sinking of science will run parallel to the sinking of many other aspects of our civilization. Indeed, they will most likely feed off each other. Science is a major characteristic of our western culture, and our way of thinking. Therefore, the end of science will mean the end of modern European culture, the twilight of an era initiated in Europe around the fifteenth century and which is extended nowadays throughout the world: the scientific age.

\section{Spengler and the Decline of the West}

Professional philosophers make very few attempts to understand the present-day problems of science, though there are some valuable rare exceptions. In most cases, these problems are only mentioned in order to discredit science in general. Why is the kind of analysis in the present article (or the full version in the book of the same title) so infrequent among the works of professional philosophers? I think there are two main reasons: 1) Most of them do not have knowledge of science at close quarters but only through reading books, which do not reflect the real problems; even in the few cases of philosophers with an education in science to the level of a scientist, they do not dedicate their time to research so they only know about present-day problems by hearsay. 2) They are not interested in revealing the miseries of another profession because they themselves share the same problems in even greater magnitude. In this panorama, what has the “office-philosopher” to say about science? Nothing different from what happens in his or her own house. At present, professional institutional philosophy has as many problems as institutional science, and reading philosophy of science written by contemporary philosophers working in a university is mostly a waste of time. Therefore, it is not strange that there should be silence about the things criticized along this article. And these problems of science are not going to go away, nor are they to be resolved by any paid philosopher (or sociologist or any group of administrators of culture paid for by the state). These questions are things to be discussed by scientists themselves, and from the inside looking out.

Nonetheless, true ``philosophers'', in the full sense of the word, even if they do not specialize in scientific questions, may offer us interesting insights about many questions related directly or indirectly with the activity of scientific research. Let us consider for instance the case of Spengler and his work {\it The Decline of the West}, a masterpiece in many ways, with plenty of lucid ideas and admirable global vision; a giant, a brave thinker with a strong character and something interesting to tell, rather than a boring treatise of trivialities and diplomatic sentences of the kind so common among our dwarf-philosophers. 

Spengler’s major conclusions are twofold: 
\begin{enumerate}
\item There are three possible ways of a civilization developing: within magical, Apollonian or Faustian conceptions of the world. Magical conceptions are associated with primitive cultures dominated by animism, religions and other superstitions, of which there are still some remnants in present-day societies. The Apollonian spirit is the characteristic of reason and logics, of the quantification of nature, typical of most scientific activities. The Faustian spirit is characterized by the will, the feeling of a direction towards an end, the consciousness of history. 

\item The second conclusion is that the history of our culture can be compared with the history of other civilizations and their developments, and in making that comparison it can be observed that western culture is declining, it is reaching the end of its possibilities. The epoch of the arts is over, and now only rubbish is being produced, as an illusion of art, keeping it alive. The epoch of philosophy is over. And the epoch of science is over too.
\end{enumerate}

\begin{quotation}
``Death of science consists of the existence of nobody able to live it. But 200 years of scientific orgies get fed up in the end. It is not the individual but the spirit of a culture who gets fed up. And this is manifest by sending to the historical world of nowadays researchers who are more and more small, mean, narrow and infecund.'' (Oswald Spengler, {\it The Decline of the West})
\end{quotation}

For Spengler, the epoch of mathematics is over, and there now remains only the work of conservation, refining, polishing, and selection. Physics is also near its limit. What remains now is an industrial production of hypotheses. Still, in the 1910s and 1920s, when Spengler wrote about this, there were important discoveries being made in physics. Nonetheless, I think Spengler is right in his prediction about the decline of the scientific world, as well as the decline of culture in general, a trend which may last many decades or centuries. In my opinion, he was ahead of his time, and he saw the problems of the future in our civilization in a prophetic way. We must bear in mind that the decadence of the Roman Empire lasted almost three centuries, after the death of Marcus Aurelius, with certain fluctuations but following a general average trend of decline (Gibbon, 1776-1789). The same thing may occur with the decline of our culture now: We have been declining all through the twentieth century, and we will continue to do it during the twenty-first century. 

\section{Conclusions}

Science is becoming a nonsense for humanity. During the last century, science has advanced more and more in technical terms, more and more in its investment in very expensive experiments, in the amount of information it generates, but it has gone backwards with regard to its motivation. The force which pushed humanity to walk towards knowledge, enlightenment and reason is now pushing very weakly. Now, science continues to work because of its inertia but is subject to some friction because to its erosion. Our science is tired, exhausted. It walks entangled with economic forces rather than with human dreams. Science has lost its first attractiveness; only simple technical operations remain.
 
Our science has become an animal without a soul, or it might be better to say, a colony of animals, a group of organisms which devour human efforts and do not offer anything but growth for the sake of growth. Scientific organizations behave like a colony of bacteria which reproduce as far as the available food/money allow. The more you feed them, the more they grow: more Ph.D. students, postdocs, staff researchers, supercomputers, telescopes, particle accelerators, papers, etc. And, if the money tap is closed, the people dedicated to science and their by-products are proportionally reduced. This is not the science of Galileo, Darwin, or Einstein, who produced their ideas when they felt the {\it spiritual} necessity to express them, independently of whether they were paid for that or not; certainly for prestige, and for the pride in revealing new truths. Nowadays, there are very few things to express; almost everything in science is reduced to find a small fiefdom of nature to analyse--whether there is any fundamental question to solve in this analysis does not matter--, and publishing papers on it and getting citations from colleagues with the aim of getting jobs and extra money for expenses. Getting money to employ more Ph.D. students, postdocs... and when these students and postdocs grow up, they become new senior researchers who ask more money, and so on. The sense of all this industry is one of primitive life: just a struggle for survival and spreading (intellectual) genes.

Why should science survive? It was important for our understanding of nature in the past, but it is not so much now. Our philosophy of nature does not change, our {\it Weltanschauung} (world view) does not change with the latest discoveries; only subtle details are now produced. Is it good for the individuals of the mankind to know so many details about nature? And if it is not produced for each of us, ``it will be, for whom?'' (Unamuno). 

The role that pure sciences (apart from technological applications) will occupy in the culture of the humanity in the future is unknown now. Let us hope it will retain our tradition of understanding how nature behaves, but in a different way from what we have known up until now. Let us wait and see if future generations keep the best of it. 

See further development of these ideas at: L\'opez Corredoira (2013).

\end{document}